# Single-step deposition of high mobility graphene at reduced temperatures

D. A. Boyd,[1] W.-H. Lin,[2] C.-C. Hsu,[1,3] M. L. Teague,[1,3] C.-C. Chen,[1,3] Y.-Y. Lo,[4] W.-Y. Chan,[5] W.-B. Su,[5] T.-C. Cheng,[4] C.-S. Chang,[5] C.-I. Wu,[4] N.-C. Yeh[1,3,6]*

[1]Department of Physics, California Institute of Technology, Pasadena, CA 91125, USA.
[2]Department of Applied Physics, California Institute of Technology, Pasadena, CA 91125, USA.
[3]Institute of Quantum Information and Matter, California Institute of Technology, Pasadena, CA 91125, USA.
[4]Graduate Institute of Photonics and Optoelectronics and Department of Electrical Engineering, National Taiwan University, Taipei, Taiwan.
[5]Institute of Physics, Academia Sinica, Nankang, Taipei, Taiwan.
[6]Kavli Nanoscience Institute, California Institute of Technology, Pasadena, CA 91125, USA.
*e-mail: Nai-Chang Yeh (ncyeh@caltech.edu)

**Current methods of chemical vapor deposition (CVD) of graphene on copper are complicated by multiple processing steps and the need for high temperatures in both preparing the copper and inducing subsequent film growth. Here we demonstrate a plasma-enhanced CVD (PECVD) chemistry that enables the entire process to take place in a single step, at reduced temperatures (< 420 °C), and in a matter of minutes. Growth on copper foils is found to nucleate from arrays of well-aligned domains, and the ensuing films possess sub-nanometer smoothness, excellent crystalline quality, low strain, few defects, and room-temperature electrical mobility up to $(6.0 \pm 1.0) \times 10^4$ cm$^2$V$^{-1}$s$^{-1}$, better than that of large, single-crystalline graphene derived from thermal CVD-growth. These results indicate that elevated temperatures and crystalline substrates are not necessary for synthesizing high-quality graphene.**

**Introduction**

Much progress has been made in growing large-area graphene by means of thermal chemical vapor deposition (CVD) based on catalytic dehydrogenation of carbon precursors on copper [1–4]. However, in many instances it is desirable to avoid multiple steps[2] and high temperatures (~1000 °C) employed in thermal-CVD growth. In particular, the conditions for the critical removal of the native copper oxide and the subsequent film growth are dissimilar enough to necessitate separate process steps. Moreover, high processing temperatures restricts the types of devices and processes where CVD can be applied and can also result in film irregularities that compromise the graphene quality itself [4]. Thermally derived strain and topological defects[5-8], for example, can induce giant pseudo-magnetic fields and charging effects, giving rise to localization and scattering of Dirac fermions[7,8] and diminishing the electrical properties.

A variant of thermal CVD, called plasma-enhanced CVD (PECVD), has been widely used for depositing many allotropes of carbon, most notably diamond films. The plasma can provide a rich chemical environment, including a mixture of radicals, molecules and ions from a simple hydrogen-hydrocarbon feedstock[9], allowing for lower deposition temperatures[10] and faster growth than thermal CVD. The ability of the plasma to concurrently support multiple reactive species is a key advantage. However, the quality of PECVD-grown graphene to date has not been significantly better than that of thermal CVD[10,11,12].

Here we demonstrate a PECVD process that encompasses both the preparation of the copper and growth of graphene in a single step. The addition of cyano radicals, which are known to etch copper at room temperature[13], to a hydrogen-methane plasma is found to produce a chemistry whereby the native oxide is removed, the copper is smoothly etched, and growth of well-aligned graphene ensues. Etching of the copper substrate is found to be self-limiting, allowing etching and growth steps to proceed in tandem under the same chemistry. The entire process occurs in a matter of minutes without active heating, and the resulting graphene films exhibit high electrical mobility and few structural defects. Our results indicate that elevated temperatures, crystalline substrates, and long process times are not necessary for synthesizing high-quality graphene.

## Results

**Description and characterization of process.** The PECVD growth process is schematically illustrated in Fig. 1a-c and further described in Methods and Supplementary Figure 1. Copper was directly exposed to a low-pressure, microwave hydrogen plasma containing trace amounts of cyano radicals (Supplementary Figure 2), methane and nitrogen (Supplementary Figure 3), as detailed in Supplementary Note 1. Removal of the native oxide and smoothing of the copper typically occurred within two minutes of igniting the plasma (Fig. 1a). Nucleation of graphene ensued on both sides of the substrate (Fig. 1b). With continued exposure to the plasma, disordered graphite and monolayer graphene covered the top and bottom sides, respectively (Fig. 1c). Copper deposits were found on both the process tube and the sample holder after successful growth runs (Fig. 1d), whereas runs with little copper etching did not produce optimal films. Optical microscopy showed that the graphene on the top surface of copper was always randomly pitted after growth (Fig. 1e), while that on the bottom surface was smooth (Fig. 1f). Scanning tunneling microscopy (STM) measurements of the surface topography of graphene on the bottom side of Cu revealed sub-nanometer flatness for samples grown on Cu foils and single crystalline Cu (100) and Cu (111), (Fig. 1g-i and Supplementary Figure 5a-c). Nitrogen incorporation into the graphene, as measured by x-ray photoemission spectroscopy (XPS), Supplementary Figure 4a-b, was found to be below the detection limit[14] of 0.1 at.%.

We found that the occurrence of PECVD-graphene growth was largely insensitive to the gas temperature of the plasma. The maximum gas temperature of the plasma determined by a thermocouple sheathed in boron nitride was found to be 160°C (425°C) for 10W (40W) plasma power, which decayed by more than 120°C (250°C) within ~1 cm from the center of the plasma. The maximum temperature of the copper substrate $T_s$ was measured using the melting points of known solids on top of a copper substrate subjected to the plasma. We found that at 40 W plasma power, lead melted whereas zinc did not, indicating that the maximum Cu substrate temperature was within the range of 327.5 °C < $T_s$ < 419.5 °C. Further, we were able to fabricate more than 300 high-quality large-area graphene samples by PECVD within 5~20 minutes in a single-step by using plasma power varying between 10 and 40 W. The typical substrate size was (8×13) mm$^2$ ~ 1 cm$^2$. Although the plasma produced by the Evenson cavity was not uniform in intensity over the length of the sample, graphene was found to cover the entire substrate and exhibited consistent quality over the entire film on the bottom side, indicating that the growth was not temperature sensitive and could occur over a range of temperatures. We expect that the sample size should be scalable in a larger cavity under the same growth conditions. The localized nature of the plasma source also allowed multiple samples to be prepared individually in the same process tube without changing conditions or breaking vacuum by simply translating the Evenson cavity.

**Nucleation and growth.** The time evolution in the Raman spectra of PECVD-graphene on opposite sides of copper foils during growth is shown in Fig. 1j-k. Spectra from the top side (Fig. 1j) featured the D-band, an indicator of edges or disorder in graphene and graphite; the G-band, a feature common to both graphene and graphite; and the 2D-band, a unique feature to graphene. The D-band was prominent throughout, while the 2D-band decreased and disappeared with time, indicating the formation of disordered graphite, which is consistent with optical images of the top side (Fig. 1e). In contrast, spectra from the bottom side (Fig.

1k) showed that the D-band decreased and eventually vanished with increasing growth time while the relative intensities of 2D and G-bands indicate the formation of monolayer graphene[15,16,17,18], as described in Methods and Supplementary Figure 6a-6b. This behavior is consistent with growth and eventual coalescence of graphene domains wherein the D-band associated with the edge states of the domains diminished[19]. A comparison of the graphene from the top and bottom sides indicate that increased mass flux and direct plasma exposure adversely affected the film quality, and hereafter we focus our studies on PECVD graphene grown on the bottom side of Cu.

Scanning electron microscopy (SEM) images taken shortly after the onset of graphene growth on copper foil (Fig. 2a-d) revealed dense, linear arrays of hexagonal graphene domains that extended across the copper grains. With increasing growth time, the well-aligned graphene domains coalesced seamlessly into a monolayer graphene with few defects. Shown in Fig. 2e-f are the 2D/G and D/G intensity ratio maps of a (100×100) μm$^2$ area of an as-deposited graphene sample on Cu, respectively. The average 2D/G intensity ratio ($I_{2D}/I_G$) is ~2 and only 5% of the mapped area show any measurable D-band. Histograms of the ($I_{2D}/I_G$) and ($I_D/I_G$) maps are provided in Supplementary Figure 6a and 6b, respectively.

Although the number of graphene layers may be determined from the full-width-half-maximum (FWHM) of the 2D-band[24], the absolute value of the FWHM can be affected by the underlying substrate[25]. To avoid such complications, Raman maps of PECVD graphene transferred from Cu to Si/SiO$_2$ are provided in Supplementary Figure 7a-7d and detailed in Methods and Supplementary Note 2. The average value of the FWHM of the 2D-band, which was fit to a single Lorentzian, was 28.8 cm$^{-1}$, and the average ratio ($I_{2D}/I_G$) = 2.7, both consistent with the figure of merit for monolayer graphene[15,16,17,18,24].

High-resolution (~ 1 nm) atomic force microscopy (AFM) and SEM studies over multiple (100×100) μm² areas of an annealed monolayer graphene sample[20], which was transferred to single crystalline sapphire with a polymer-free technique[21], indicated no discernible grain boundaries. With an increase of growth time, small adlayers began to develop on top of the first layer in an aligned fashion, as exemplified by the SEM images in Fig. 2h-j. These findings, which are supported by Fig. 1k, indicate that the growth of PECVD-graphene on copper foils proceeded by nucleation and growth of well-aligned domains[22,23] that eventually coalesced into a large single crystalline sheet, and the subsequent growth of the second layer with increasing time followed a similar mechanism (Fig. 2h-j).

**Strain and structural ordering.** The biaxial strain $\left(\varepsilon_{ll}+\varepsilon_{tt}\right)$ in the PECVD-graphene on Cu can be estimated by considering the Raman frequency shifts $\Delta\omega_m \equiv \left(\omega_m - \omega_m^0\right)$ and the Grüneisen parameter $\gamma_m^{biax}$:[15,26]

$$\gamma_m^{biax} = \frac{\Delta\omega_m}{\omega_m^0\left(\varepsilon_{ll}+\varepsilon_{tt}\right)}, \tag{1}$$

where m (= G, 2D) refers to the specific Raman mode. Using the parameter $\gamma_{2D}^{biax} = 2.7$, the strain distribution over an area of (100×100) μm² is exemplified in Fig. 2g, showing consistently low strain characteristics, with an average strain ~ 0.07% for m = 2D.

Overall, a general trend of downshifted G-band and 2D-band[15] was found for all PECVD-graphene relative to thermal CVD-grown graphene on the same substrate, as exemplified in Supplementary Figure 8a-c for comparison of PECVD- and as-grown thermal CVD-graphene on substrates of Cu foil, Cu (100) and Cu (111). The consistent frequency downshifts for all PECVD-graphene indicate reduction in the averaged biaxial strain. Further, absence of the D-band in most spectra suggests that the samples were largely free of disorder/edges on the macroscopic scale, which is further corroborated by detailed spatial mapping of the Raman spectra over an area of (100×100) μm² in Fig. 2e-f.

Our PECVD-graphene samples also exhibited well-ordered honeycomb atomic lattices, which is unique to monolayer graphene, as evidenced by the STM images of samples on Cu foil (Fig. 3a-c), Cu (100) (Fig. 3e-g), and Cu (111) (Fig. 3i-k). The long-range structural ordering was further corroborated by the sharp Fourier transformation (FT) of STM topographies, as exemplified in Fig. 3d,h,l. The FT spectra demonstrated dominantly hexagonal lattices for all PECVD-graphene. Samples grown on Cu single crystals further exhibited Moiré patterns, as manifested by the second set of smaller Bragg diffraction peaks. Simulations (see Methods) indicated the Moiré pattern (a parallelogram) in the FT of Fig. 3h for graphene on Cu (100), was the result of a square lattice at approximately an angle ~ (12±2)° relative to the honeycomb lattice (Supplementary Figure 9a), whereas the Moiré pattern (a smaller hexagon) in the FT of Fig. 3l suggested that the Cu (111) hexagonal lattice was at ~ (6±2)° relative to the graphene lattice (Supplementary Figure 9b). These STM measurement over multiple samples and multiple areas per sample confirm the predominance of monolayer graphene and the absence of discernible bilayer graphene in our PECVD-grown samples.

The STM topography was further employed to analyze strain at the microscopic scale. For a local two-dimensional displacement field $\mathbf{u} = u_x\hat{x} + u_y\hat{y} \equiv \mathbf{r} - \mathbf{r}_0$, where r and r₀ denote the actual position of a carbon atom and its equilibrium position in ideal graphene, respectively[27], the compression/dilation strain is given by $(\partial u_x/\partial x) + (\partial u_y/\partial y) \equiv u_{xx} + u_{yy}$, which is proportional to the biaxial strain[5,7,8]. Using the topographies shown in Fig. 3a-c, we obtained the spatial strain maps for PECVD-graphene on Cu foil, Cu (100) and Cu (111) substrates over successively decreasing areas in Fig. 4a-b, 4e-f and 4i-j, respectively. The corresponding strain histograms are given in Fig. 4c,g,k. The PECVD-graphene exhibited low and relatively homogeneous strain distributions. Further comparison with the macroscopic strain obtained from a collection of Raman spectra taken on different areas of multiple PECVD-graphene samples are summarized by the strain histograms in Fig. 4d,h,l. There is overall consistency between microscopic STM studies and macroscopic Raman spectroscopic studies, revealing low strain for all PECVD-graphene.

**Electrical properties.** The electrical mobility ($\mu$) of PECVD-graphene was determined by studying back-gated field-effect-transistor (FET) devices[28,29]. Graphene samples were first transferred to hexagonal BN thin films on 300 nm $SiO_2$/Si substrates using a polymer-free method[21], and then lithographically processed into a geometry shown in Fig. 5a-b. The sheet resistance of graphene as a function of the back-gate voltage ($V$) and the corresponding conductivity ($\sigma$) vs. sheet carrier density ($n_s$) were measured at 300 K, as exemplified in Fig. 5c-d. The electrical mobility $\mu$ was obtained from the derivative of the Drude formula[28] near the charge neutrality point:

$$\mu = (1/C) d\sigma/dV , \qquad (2)$$

where $C$ denotes the capacitance of the device. The electrical mobility obtained from 9 different devices was found to range from $(3.0\pm0.5)\times10^4$ $cm^2V^{-1}s^{-1}$ to $(6.0\pm1.0)\times10^4$ $cm^2V^{-1}s^{-1}$ for electrons, as summarized in Table 1, and from $(1.2\pm0.2)\times10^4$ $cm^2V^{-1}s^{-1}$ to $(3.5\pm0.5)\times10^4$ $cm^2V^{-1}s^{-1}$ for holes, where the errors represented variations in determining the $(d\sigma/dV)$ slope due to the discreteness of voltage readings. These values are comparable to those obtained on large, single crystalline thermal CVD-grown graphene on BN, the latter yielded $\mu \sim 4\times10^4$ to $6\times10^4$ $cm^2V^{-1}s^{-1}$ at 1.7 K and $\sim 1.5\times10^4$ to $3\times10^4$ $cm^2V^{-1}s^{-1}$ at 300 K.[2]

It is worth commenting on the accuracy of mobility values obtained by the two-terminal configuration depicted in Fig. 5a-b. The contact resistance between electrode and graphene was determined by the Transmission Line Models (TLM) method and was found to be typically ~ 10 Ω for all devices. In contrast, the sheet resistance of graphene was typically more than 1 kΩ near the charge neutrality point, which was about 2 orders of magnitude larger than the contact resistance. Therefore, the contact resistance would not affect the accuracy of the mobility significantly. Additional measurements by patterning some of the PECVD-graphene into the four-point configuration revealed that the differences in mobility thus determined was less than 2-3 % from those obtained by means of the two-point method.

**Discussion**

Experience with diamond PECVD lends understanding to the observed nucleation and growth of aligned graphene domains presented here. In the case of diamond growth, it is known to involve a competition between growth by carbon radicals (most notably methyl radicals) and etching of amorphous or disordered carbon by atomic hydrogen. The process conditions used in our experiments are similar to those of microwave CVD growth of diamond thin films. Therefore, it is not unreasonable to expect that our PECVD growth of graphene also proceeds in a competitive manner.[30] Further, diamond is known to preferentially nucleate via surface defects[31,32], and the aligned domains in our work were found to be commensurate with the inherent machine marks on the as-received copper foils. Although the copper surface is smoothed during the PECVD process, remnants of the defects are likely because the process temperature is low. For comparison, aligned nucleation was not observed on single-crystal copper, which contains no marks. For thermal CVD growth of large, single-crystal graphene[2], it is preferable to have smooth, defect-free substrates and restricted nucleation. Therefore, multiple steps to prepare the substrates prior to thermal CVD growth and post annealing after the growth to optimize the sample are necessary[2]. In contrast, graphene with equivalent or even better mobility can be achieved by the one-step PECVD process described in this work. This PECVD process is also scalable[11], occurs at CMOS compatible temperatures, and avoids complications with multiple steps and post-processing. Overall, our findings of the guided PECVD growth process not only shed new light on the growth kinetics of graphene but also open up a new pathway to large-scale, excellent quality and fast graphene fabrication. In particular, this one-step PECVD process potentially allows graphene to be used as-deposited, making it amenable for integration with complementary materials and technology.

## Methods

**Experimental setup.** The experimental setup is summarized in Supplementary Figure 1, which consists of plasma, vacuum, and gas delivery systems. The plasma system (Opthos Instruments Inc.) consists of an Evenson cavity and a power supply (MPG-4), which provides an exciting frequency of 2450 MHz. The Evenson cavity mates with a quartz tube of inner and outer diameters of 10 mm and 12.5 mm, respectively. The vacuum system is comprised of a mechanical roughing pump, two capacitance manometers, a pressure control valve, and a measurement control system. There is a fore-line trap between the vacuum pump and the MKS 153 control valve. The gas delivery system consists of mass flow controllers (MFCs) for $H_2$, $CH_4$, and Ar. A bakeable variable leak valve is placed before the methane MFC, and there is a leak valve for $N_2$. A gas purifier was place before the variable leak valve on the methane line. Quarter-turn, shut-off valves are placed directly after each of the MFCs. The system pressure and gas flows are monitored and controlled through the controller via a LabView interface. A residual gas analyzer (RGA) is used to monitor the exhaust gas.

**Detailed PECVD growth procedure and information.** The copper substrates were placed on a quartz flat inside of quartz tube. A typical substrate size was (8×13) mm$^2$. The tube was evacuated to 25 to 30 mTorr. A 2-5 sccm flow of room temperature hydrogen gas with 0.4% methane and a comparable amount of nitrogen gas was added and the pressure was controlled at 500 mTorr.

The addition of methane to the gas flow was controlled by a precision leak valve, and a typical concentration, as measured by gas chromatography, was 0.4 %. The initially low vacuum conditions (~ 25 mTorr) provided sufficient partial pressures of nitrogen (atmospheric) for the PECVD process. (We note that a leak valve for including purified nitrogen was added to the setup to study the effects of increased nitrogen partial pressures.) Typical partial pressures measured by residual gas analysis (RGA) are as follows:

    $N_2$ (Mass 28):    $6.9 \times 10^{-8}$ Torr
    $CH_4$ (Mass 16):    $1.5 \times 10^{-7}$ Torr

| | |
|---|---|
| O$_2$ (Mass 32): | 5.4 × 10$^{-9}$ Torr |
| H$_2$O (Mass 18): | 1.2 × 10$^{-6}$ Torr |
| CO$_2$ (Mass 44): | 2.3 × 10$^{-8}$ Torr |

From these values we estimate that the concentration of nitrogen in the gas flow is typically on the same order as that of methane.

The PECVD process was found to be highly sensitive to the relative amounts of methane and nitrogen. Excessive methane mixtures resulted in no etching, while excessive nitrogen mixtures would result in excessive etching of the copper. Slight adjustments to the methane concentration could be made via either optical emission spectroscopy (OES) or residual gas analysis (RGA).

A low-power (ranging from 10 to 40 W) cold hydrogen plasma was formed over the copper substrate using an Evenson cavity. Exposure of copper substrates to the plasma enabled continuing etching and cleansing of the copper surface during the graphene growth process. The gas temperature inside the plasma was measured using a thermocouple sheathed in boron nitride. The thermocouple was placed in the plasma above the sample and could be translated along the tube. The peak gas temperature measured in plasma treatment was 160 °C at 10 W and 425 °C at 40 W, and the gas temperature profile decreased rapidly (by 120 °C at 10 W and 425 °C at 40 W within 1 cm) from the peak value. The maximum temperature of the copper substrate $T_s$ was measured using the melting point of known solids, lead and zinc, and found to be within the range of 327.5 °C < $T_s$ < 419.5 °C at 40 W. Typically after 5 to 20 minutes of direct exposure to the plasma, a large-area monolayer graphene formed on the backside of copper substrates while the front side of the substrates was coated with disordered graphite. Upon the completion of graphene growth, the plasma was extinguished, and the gas flows were stopped. The process tube was then evacuated and back filled with Ar, and the substrate with graphene coating was subsequently removed. Copper deposition was visible on the inside of the tube and on the sample holder as the result of plasma etching of the copper substrates.

**Analysis of monolayer graphene using Raman spectroscopy.** The Raman maps shown in Fig. 2e-f of the manuscript for PECVD-graphene on Cu were collected with Renishaw InVia @ 532 nm, and the spectra were taken at 2 μm per pixel steps over an area of (100 × 100) μm$^2$ for a total of 2601 spectra. Similarly, the Raman maps in Supplementary figure 7a-b for PECVD-graphene transferred to SiO$_2$ were taken at 2 μm per pixel steps over an area of (160 × 150) μm$^2$. Each spectrum was smoothed and the broad background from the copper fluorescence was removed, and the peak locations and intensities for the D, G, and 2D features were extracted using a Matlab script. The relative intensities of the 2D and G-band and those of the D and G-band are presented Fig. 2e-f, and histograms of the ratio values and the FWHM linewidth of the 2D band of our PECVD-grown graphene on Cu and commercial thermal CVD-grown graphene on Cu are shown in Supplementary figure 6a-e. The predominant ($I_D/I_G$) ~ 0 value of the graphene sample shown in Supplementary figure 6a indicates negligible defects, whereas the histogram of the 2D/G intensity ratio ($I_{2D}/I_G$) in Supplementary figure 6b reveals that the average of the ($I_{2D}/I_G$) value is ~ 2, which agrees with the figure of merit ($I_{2D}/I_G$) > 1 for monolayer graphene[15,16,17,18].

This notion of predominantly monolayer graphene described in this work is further corroborated by detailed studies of a PECVD-graphene sample transferred from Cu to a SiO$_2$ substrate, which prevents the complication of slow oxidation of the Cu substrate that could result in graphene linewidth broadening with time due to the influence of the substrate (see Supplementary Note 2). As shown in Supplementary Figure 7a-d, a large spatial map over an area of (160×150) μm$^2$ revealed a mean FWHM value = 28.8 cm$^{-1}$ for the 2D-band and a mean value ($I_{2D}/I_G$) = 2.7, both are consistent with predominantly monolayer graphene if we use either the criterion of 2D-band FWHM < 30 cm$^{-1}$ as the figure of merit for monolayer graphene on SiO$_2$,[24] or the criterion ($I_{2D}/I_G$) > 1 for monolayer graphene on various substrates[15,16,17,18].

**Simulations.** Simulations of the Moiré patterns and the corresponding Fourier transformation (FT) were made by means of Matlab. We generated the triangular and cube lattices by using cosine square function and the hexagonal lattice by displacement of two triangular lattice functions. The lattice constant for the ideal 2D honeycomb structure of graphene is 0.2461 nm and that for the ideal Cu FCC lattice is 0.3615 nm, which has a corresponding lattice constant $(0.3615/\sqrt{2}) = 0.2556$ nm for the triangular lattice in the 111 direction and the square lattice in the 100 direction.

To simulate realistic STM imaging of the Moiré pattern for a given relative angle $\theta$ between the top graphene layer and the underlying Cu lattice, we assume the resulted topological function $G(x,y)$ has the following form:

$$G(x,y) = G_C(x,y) + r\, e^{-k G_C(x,y)} G_{Cu}(x,y,\theta) , \qquad (3)$$

where $0 < r \leq 1$, $G_C(x,y)$ is the graphene honeycomb lattice function and $G_{Cu}(x,y,\theta)$ is the copper layer function which could be either the triangular lattice for Cu (111) or the cubic lattice for Cu (100). Both functions are normalized with maximum 1 and minimum 0. We use the exponent $e^{-k G_C(x,y)}$ in equation (3) to simulate the rapid decay of the tunneling current contribution from the copper atom directly under a carbon atom in the top graphene layer, where $k$ is parameter that controls the decay rate and was chosen to be 2 ~ 4. Even for exposed copper atoms without any carbon atom directly positioning above, the signal from copper is expected to be attenuated by a ratio $r$ due to a larger distance from the STM tip to the copper layer. The above expression ensures that contributions from copper atoms directly under carbon atoms are much reduced due to deflection of tunneling electrons, whereas those from exposed copper atoms are only attenuated by a coefficient $r$.

The matching between simulations and the Fourier transform (FT) of graphene on Cu (100) lattice data is judged by the angle of inner parallelism. At $\theta = (12\pm2)°$, the angle of the inner parallelism appear to match the experimental result in Fig. 3h, as shown in by the real space Moiré pattern (left panel) and the corresponding FT (right panel) in Supplementary Figure 9a.

In contrast, the matching between simulations and the FT of graphene on Cu (111) data is based on the size of inner hexagon relative to that of the outer hexagon for graphene lattice. In fact, there are actually two inner hexagons according to the simulations, as exemplified in the FT of the Moiré pattern in Supplementary Figure 9b for $\theta = (6\pm2)°$. However, empirically only the larger inner hexagon is visible (Fig. 3l). We attribute this discrepancy to the intense DC signal from the zone center of the FT spectra, which overwhelms the signal of the smaller inner hexagon and so becomes invisible upon our removal of the DC signal.

## Acknowledgements

This work at Caltech was supported by National Science Foundation under the Institute of Quantum Information and Matter, and by Moore and Kavli Foundations through the Kavli Nanoscience Institute. The work in Taiwan was supported by the National Science Council under contracts 100-2911-I-002-514 and 101-2628-M-002-004. The authors thank Professor George Rossman for the use of his Raman spectrometer. DAB specially thanks late Professor David G. Goodwin for his friendship and mentorship.

## Author contributions

D.A.B. conceived the PECVD graphene growth idea. D.A.B., W.-H.L. and C.-C.H. developed the PECVD graphene growth procedures and carried out Raman spectroscopic characterizations. C.-C.H. carried out SEM and AFM studies on graphene samples. W.-H.L., Y.-Y.L. and C.-I.W. processed the graphene FET devices and carried out the mobility measurements and analysis. W.-B.S., C.-S.C. and M.L.T. performed the STM studies. C.-C.C. and M.L.T. conducted the simulations and analysis of the STM data. T.-C. C. performed the XPS studies and analysis of the graphene samples. N.-C.Y. coordinated with all co-authors on the design, planning and execution of the experiments, data analysis and simulations, and wrote the manuscript together with D.A.B.

## Additional information

Supplementary information is available in the online version of the paper. Reprints and permissions information is available online at www.nature.com/reprints.

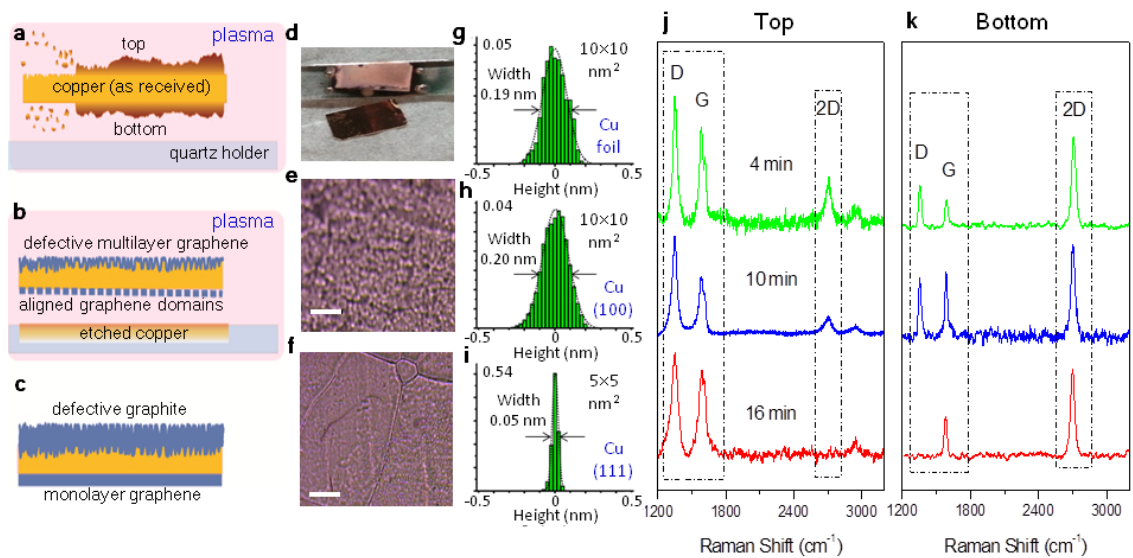

**Figure 1 | PECVD-graphene growth process:** (**a**) Exposure to plasma removes the native oxide and smoothes the copper substrate; (**b**) Aligned graphene nucleates on the bottom of the copper and disordered multilayer graphene forms on the top; (**c**) Monolayer graphene and disordered carbon develop respectively on the bottom and top of the copper substrate with continued exposure to plasma. (**d**) A copper foil and the sample holder, showing etched copper after PECVD growth. Optical images of the top (**e**) and bottom (**f**) of a copper foil after growth, where the scale bars correspond to 50 μm. (**g**) Height histogram for PECVD-graphene grown on Cu foil. (**h**) Height histogram for PECVD-graphene grown on single crystalline Cu (100). (**i**) Height histogram for PECVD-graphene grown on Cu (111). (**j-k**) Comparison of the time-evolved Raman spectra of the (**j**) top and (**k**) bottom of the Cu foil taken with increasing growth time.

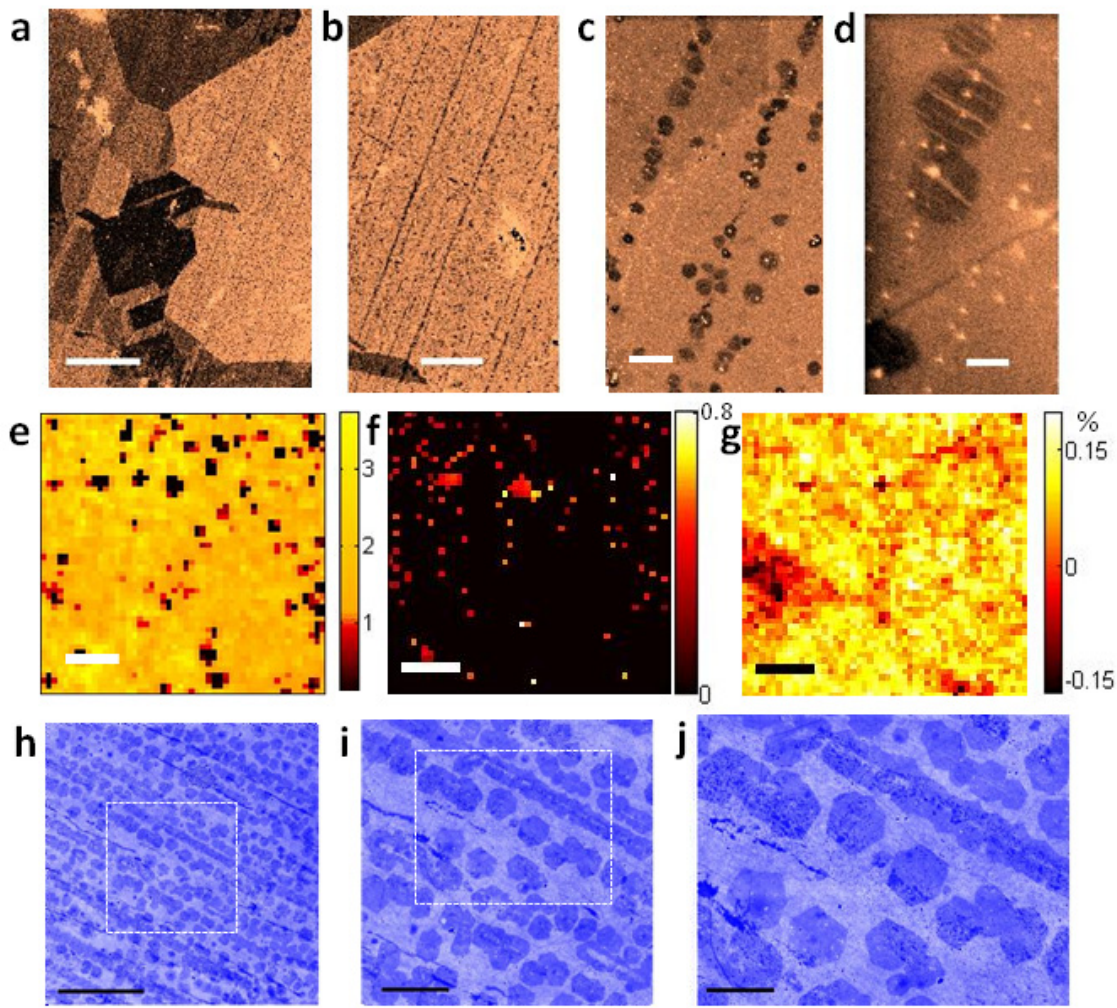

**Figure 2 | Large-area characterization of PECVD-graphene on Cu:** (**a-d**) False-colour SEM images of early-stage growth (with increasing magnification from left to right, where the scale bars correspond to 30 μm, 10 μm, 1 μm and 200 nm, respectively), showing extended linear arrays of well-aligned hexagonal domains (dark) on copper foil (light); (**e**) (100×100) μm$^2$ map (scale bar: 20 μm) of the Raman spectral 2D/G intensity ratio of a fully developed monolayer graphene sample on copper foil; (**f**) (100×100) μm$^2$ map (scale bar: 20 μm) of the Raman spectral D/G intensity ratio over the same area as in **e**; (**g**) (100×100) μm$^2$ strain map (scale bar: 20 μm) over the same area as in **e** and **f**; (**h-j**) False-colour SEM images of graphene grown for excessive time and transferred to single crystalline sapphire (with increasing magnification from left to right, where the scale bars corresponding to 30 μm, 10 μm and 5 μm, respectively), showing well aligned adlayer graphene domains (dark) on the bottom monolayer graphene (light).

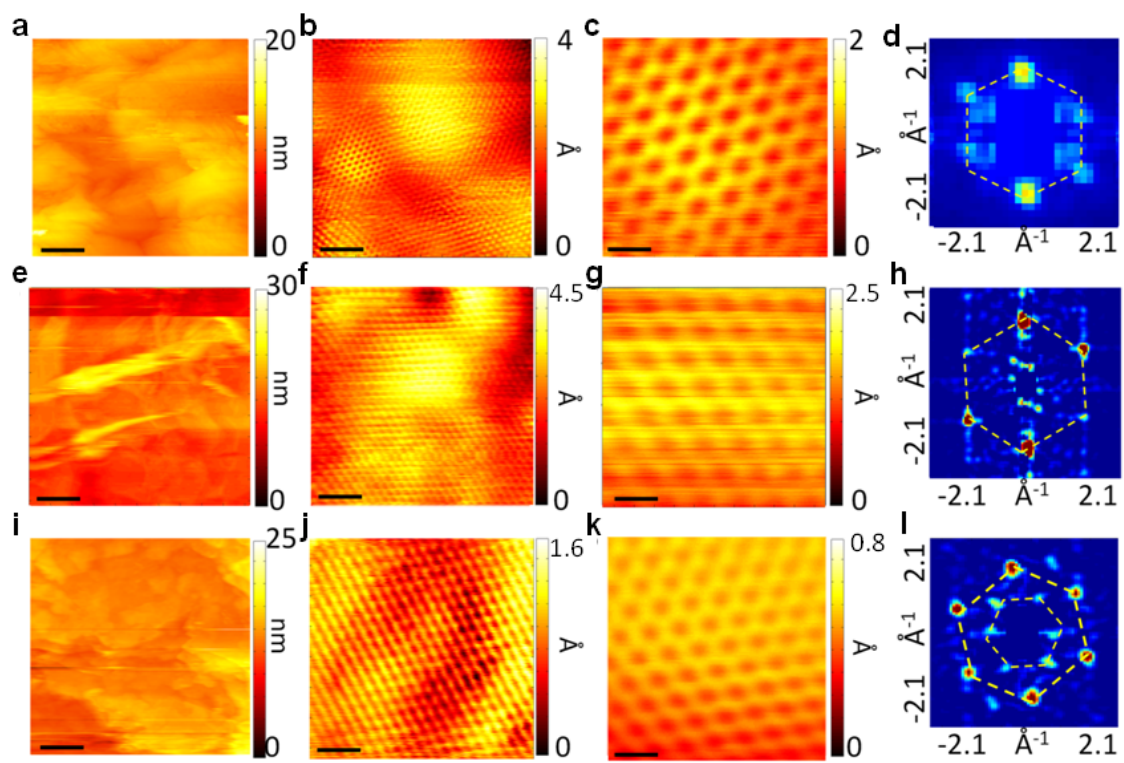

**Figure 3 | Topographies of PECVD-graphene on copper:** STM topographies of PECVD graphene at 77 K over successively decreasing areas (first three columns) and the corresponding Fourier transformation (FT) of large-area topography (fourth column) for samples grown on (**a-d**) Cu foil; (**e-h**) Cu (100); and (**i-l**) Cu (111). The scale bars for **a**, **e** and **i** are 40 nm; for **b**, **f** and **j** are 2 nm; and for **c**, **g** and **k** are 0.4 nm.

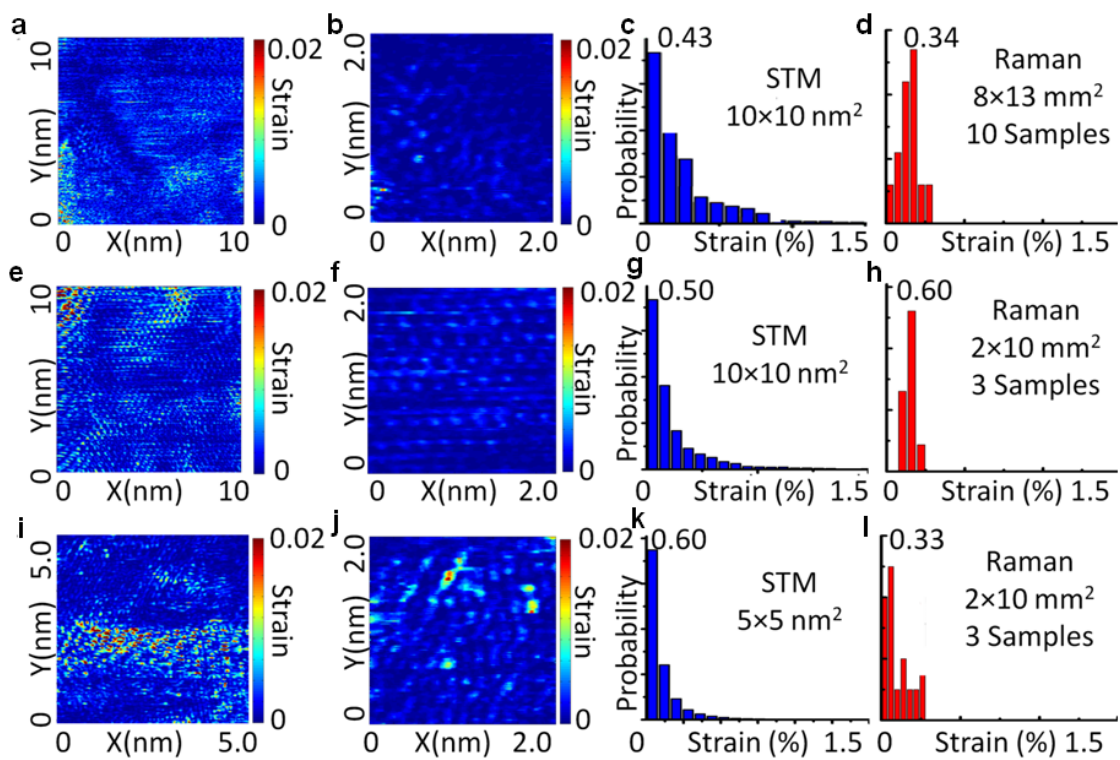

**Figure 4 | Comparison of the spatially resolved strain maps and strain histograms of PECVD-graphene:** From left to right, compression/dilation strain maps over successively decreasing areas taken with STM at 77 K (first and second columns, colour scale in units of %), strain histogram (third column) of the strain map shown in the first column, and strain histogram (fourth column) obtained from Raman spectroscopic studies of different areas of multiple PECVD-graphene samples grown on (**a-d**) Cu foil; (**e-h**) Cu (100); and (**i-l**) Cu (111). The strain obtained from STM topography is largely consistent with the findings from Raman spectroscopic studies.

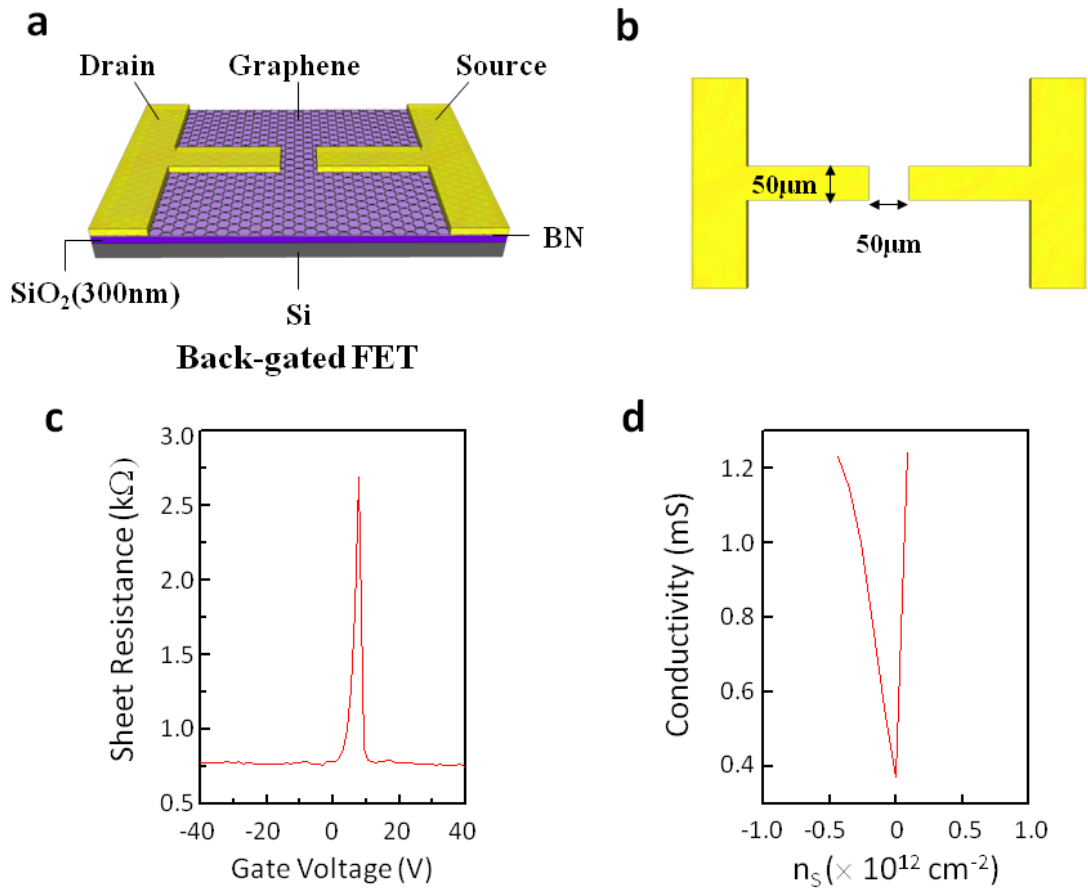

**Figure 5 | Electrical mobility measurements:** (**a**) Schematics of a typical back-gated FET device and (**b**) detailed dimensions of the top view; (**c**) Representative sheet resistance vs. gate voltage ($V$) data taken at 300 K for PECVD-graphene on Cu foil and transferred to BN; (**d**) Conductivity ($\sigma$) vs. sheet carrier density ($n_s$) data converted from a. A set of electron mobility ($\mu$) data from 9 different FET devices based on PECVD-graphene grown on Cu foils are summarized in Table 1, where $\mu = (1/C)(d\sigma/dV)$ was determined at the charge neutrality point. The errors for each device due to discreteness of the voltage readings exemplified in c are within ±15% of the tabulated mean value.

**Table 1** | Electron mobility ($\mu$) data from 9 different FET devices based on PECVD-graphene

| Device | Electron Mobility (cm$^2$V$^{-1}$s$^{-1}$) |
|:---:|:---:|
| 1 | 54,000 |
| 2 | 52,000 |
| 3 | 42,000 |
| 4 | 38,000 |
| 5 | 31,000 |
| 6 | 30,000 |
| 7 | 55,000 |
| 8 | 60,000 |
| 9 | 38,000 |
| Average | 44,400 |

**Supplementary Information**

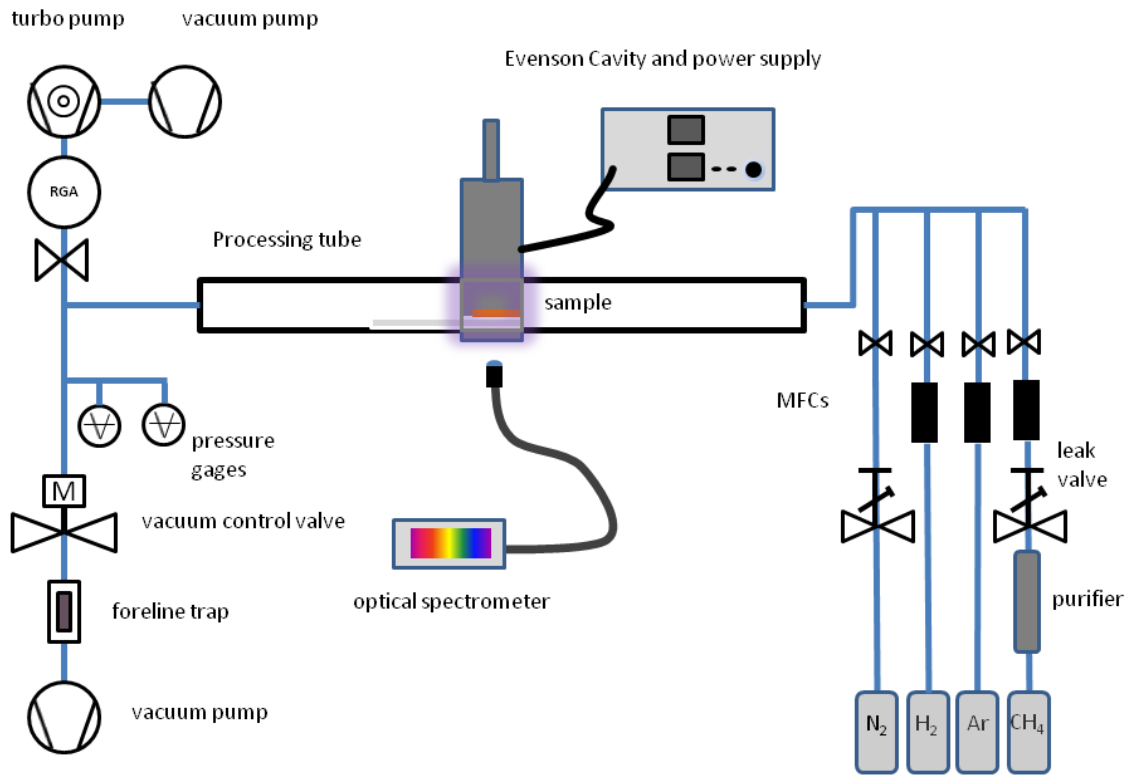

**Supplementary Figure 1.** A schematic of the experimental setup used for graphene fabrication.

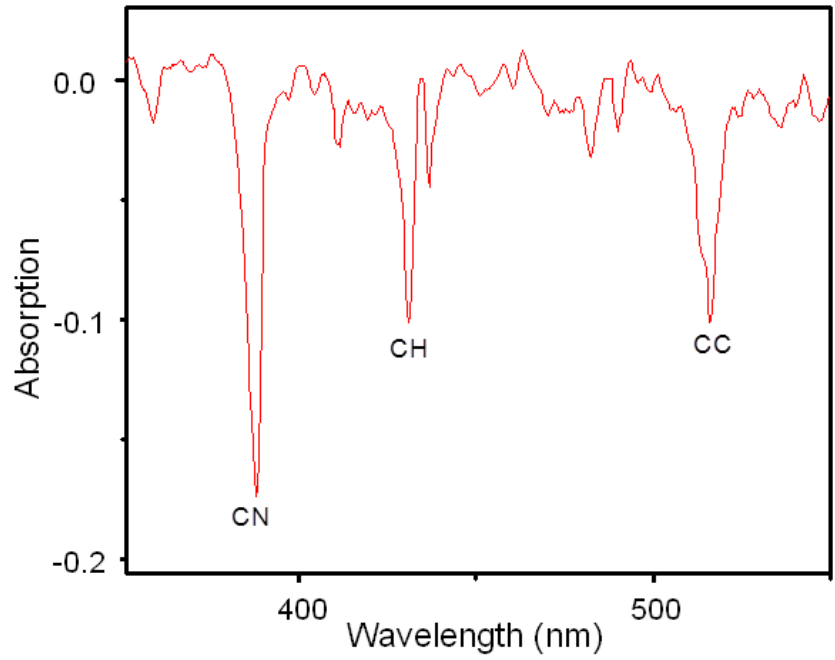

**Supplementary Figure 2. Emission spectrum of the plasma:** Negative peaks indicate an increase in particular species while positive peaks indicate a decrease.

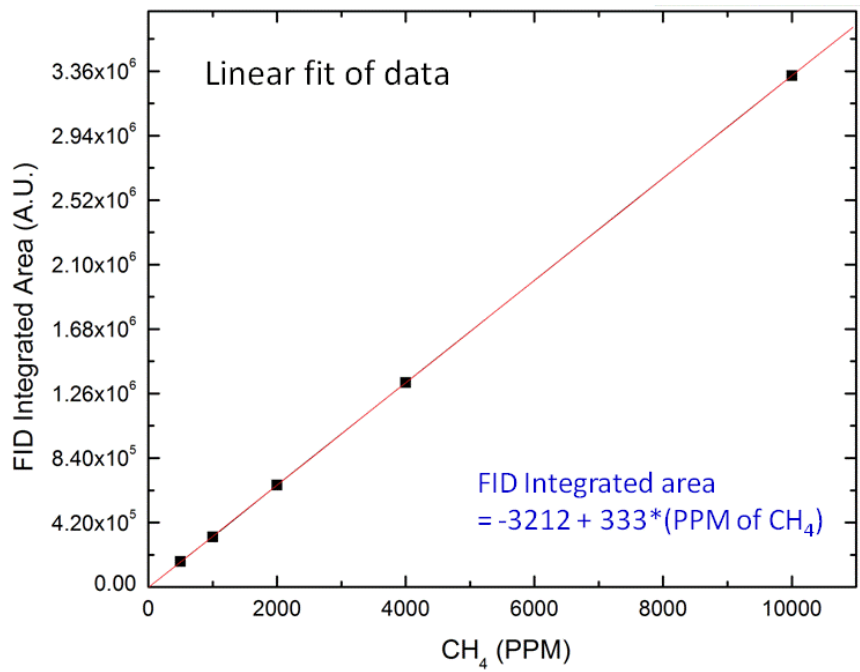

**Supplementary Figure 3. Response of the FID detector to methane in nitrogen:** The data are shown in units of PPM, and the line is a linear fit of the FID integrated area.

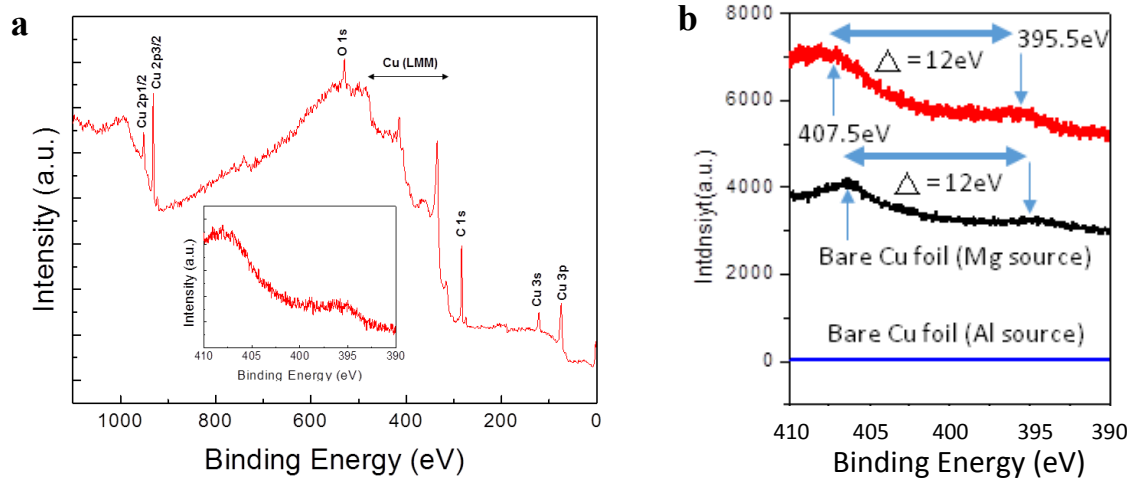

**Supplementary Figure 4. XPS studies of PECVD-graphene on Cu:** (**a**) XPS spectrum of a PECVD-graphene sample on Cu foil, showing no discernible signals at the N 1s core level binding energy (398.1eV). (**b**) Further investigation of two small peaks at 407.5 eV (Peak A) and 395.5 eV (Peak B) are attributed to the Auger lines of underlying Cu, as manifested by the consistent XPS spectral features for a bare Cu foil (black curve) with those of our graphene sample on Cu (red curve). The shifting in energy of the two peaks with changing the source of XPS from Mg to Al (blue curve) further support the fact that the peaks are from the Auger lines of Cu rather than the nitrogen 1s core level because the binding energy of nitrogen is independent of the x-ray energy.

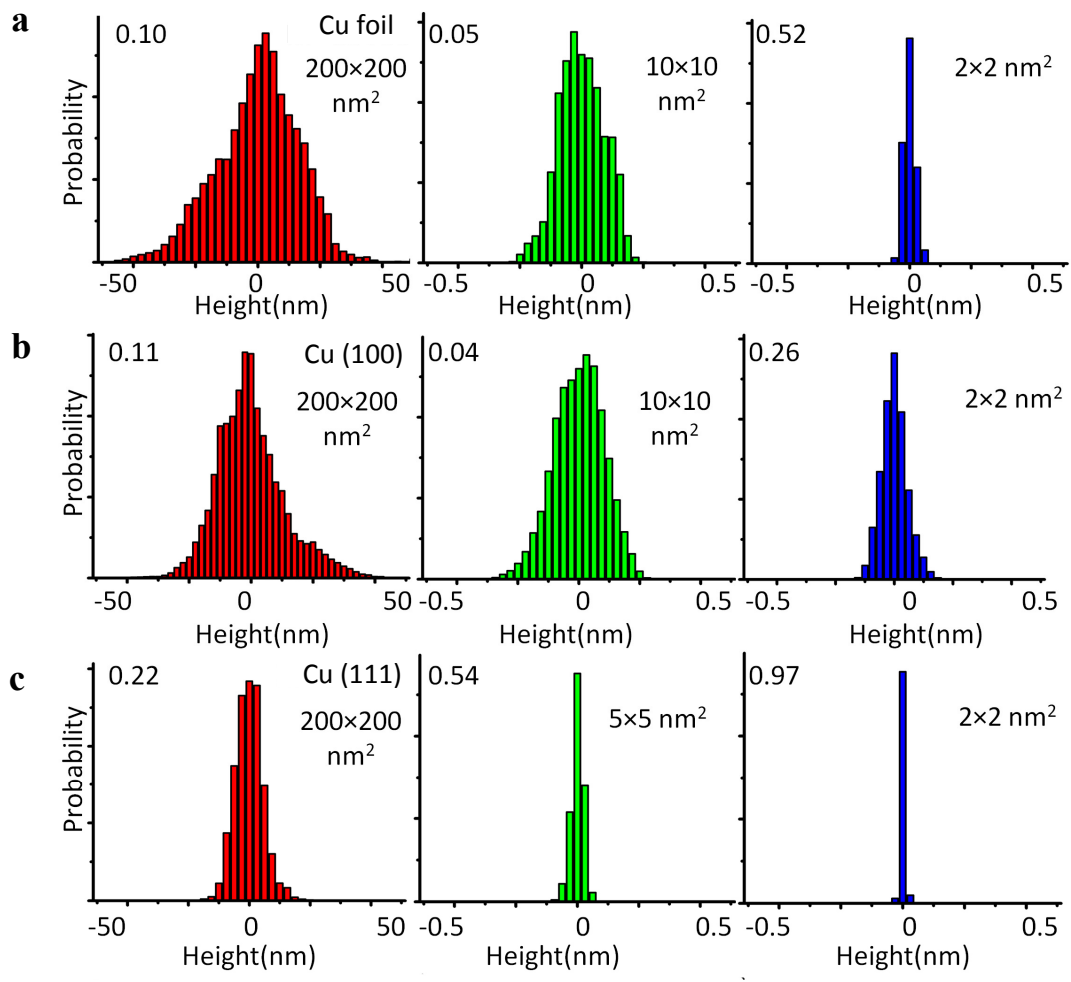

**Supplementary Figure 5. STM studies of the surface morphology of PECVD-graphene on different substrates:** From left to right, the height histograms of successively decreasing areas for PECVD-graphene on (**a**) Cu foil; (**b**) Cu (100) single crystal; and (**c**) Cu (111) single crystal. The overall surface morphology for PECVD-grown graphene appears to be much smoother than that of the 1000 °C thermal CVD-grown graphene at all length scales.

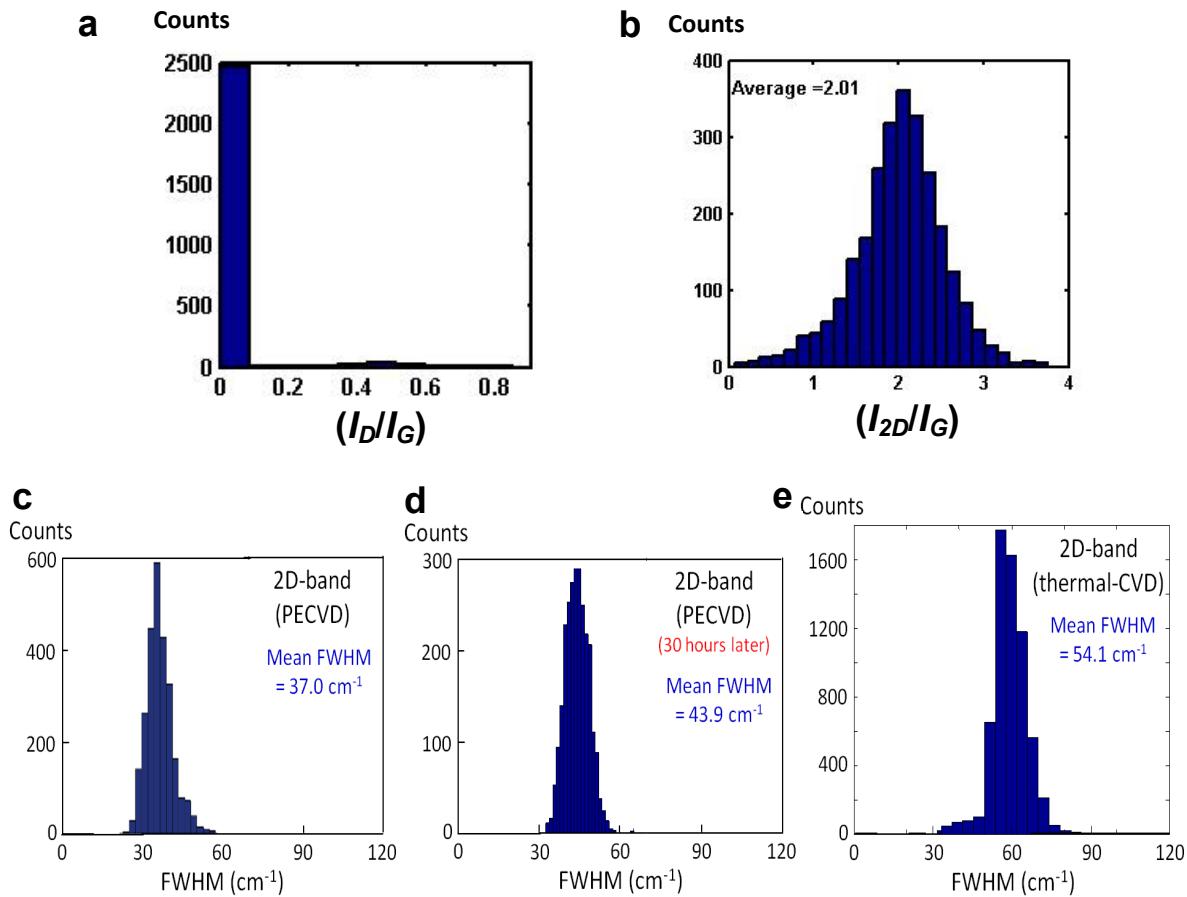

**Supplementary Figure 6. Raman spectral analysis of PECVD-graphene on Cu:** (a) Histogram of the *D*/*G* intensity ratios ($I_D/I_G$) of the Raman map in Figure 2b, showing nearly complete absence of the *D*-band signal (which is associated with defects) in the PECVD-grown graphene. (b) Histogram of the 2D/*G* intensity ratios ($I_{2D}/I_G$) of the Raman map in Figure 2b, showing the majority of the ratios exceeding 1, which is a figure of merit for the quality of monolayer graphene. (c) Histogram of the 2D-band linewidth (FWHM) of the same PECVD-graphene on Cu taken several days after growth, showing a mean FWHM = 37.0 cm$^{-1}$. (d) Histogram of the 2D-band FWHM of the same sample as in (c) taken at 30 hours later while continuously exposed to atmosphere, showing a mean FWHM = 43.9 cm$^{-1}$. (e) A comparative histogram of the 2D-band FWHM of a commercially purchased thermal-CVD grown graphene on Cu (Graphene Supermarket), showing a mean FWHM = 54.1 cm$^{-1}$.

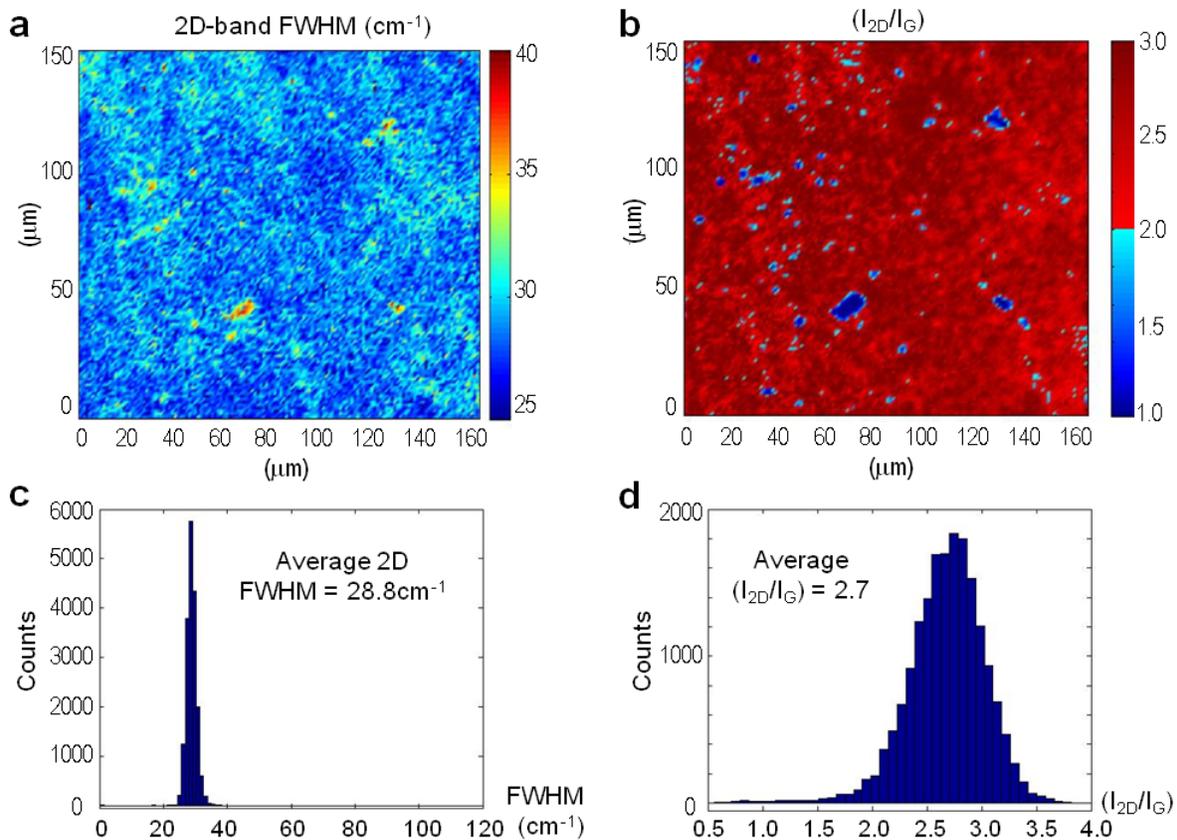

**Supplementary Figure 7. Raman spectral analysis of PECVD-graphene transferred from Cu to SiO$_2$ after one year of its growth:** (a) Spatial map of the FWHM of the 2D-band over an area of (160×150) µm$^2$, showing typical FWHM values < 30 cm$^{-1}$, which is a figure of merit for monolayer graphene according to S. Lee *et al.*, Nano Lett. **10,** 4702 (2010). (b) Spatial map of the 2D/*G* intensity ratios ($I_{2D}/I_G$) over the same sample area of as in (a), showing ($I_{2D}/I_G$) > 2 over most of the sample, again consistent with the figure of merit ($I_{2D}/I_G$) > 1 for monolayer graphene. (c) Histogram of the FWHM of the 2D-band in (a), showing a mean FWHM = 28.8 cm$^{-1}$. (d) Histogram of the 2D/*G* intensity ratios ($I_{2D}/I_G$) in (b), showing a mean value of ($I_{2D}/I_G$) = 2.7.

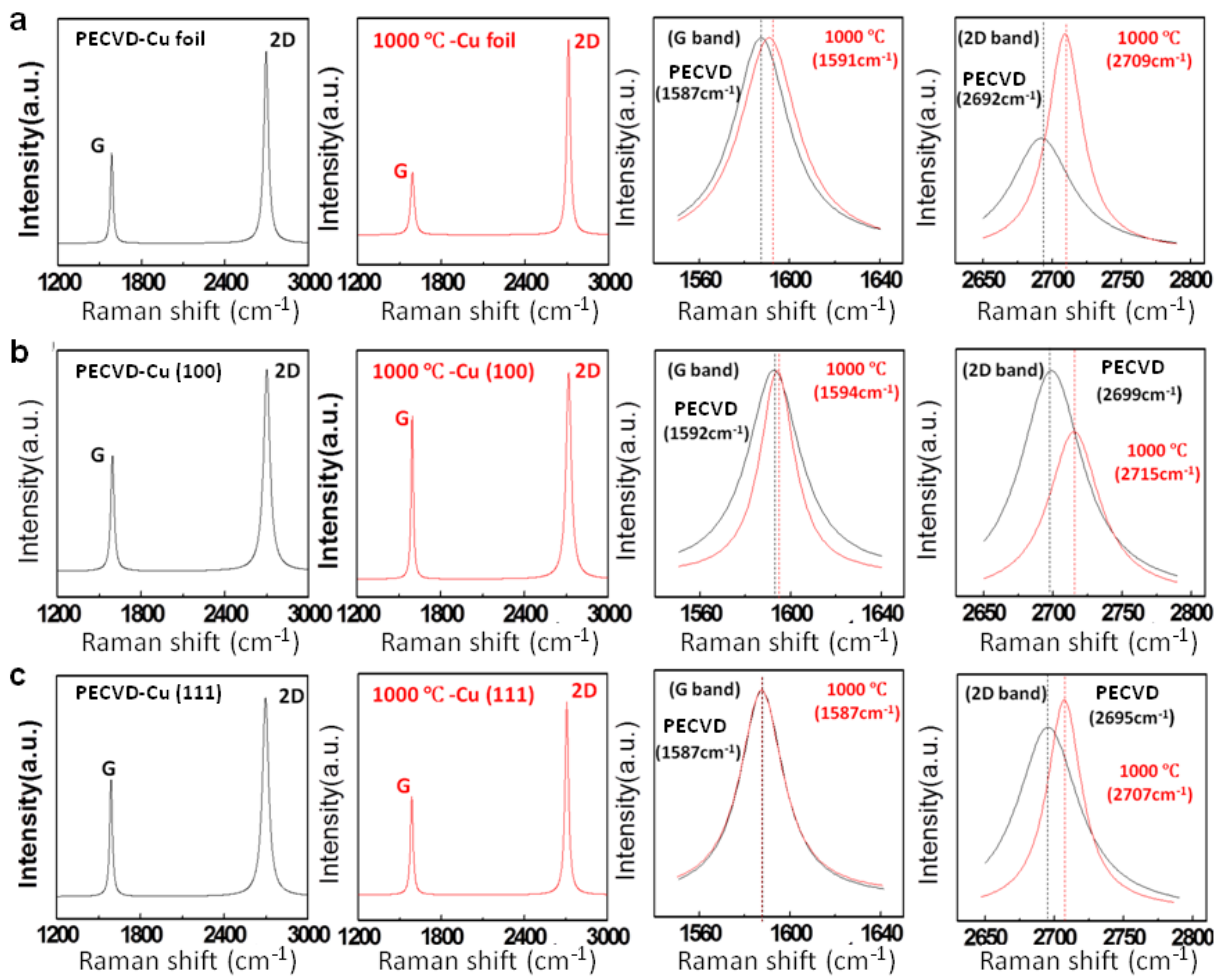

**Supplementary Figure 8. Comparison of the Raman spectroscopy of PECVD-graphene and thermal CVD-grown graphene on different substrates:** From left to right, Raman spectra of PECVD-graphene, 1000 °C-grown graphene, comparison of the 2D-band peaks, and comparison of the zone-centre G-band peaks for samples grown on (a) Cu foils, (b) Cu (100) and (c) Cu (111). The Raman spectra taken here were collected with a Renishaw M1000 @ 514 nm.

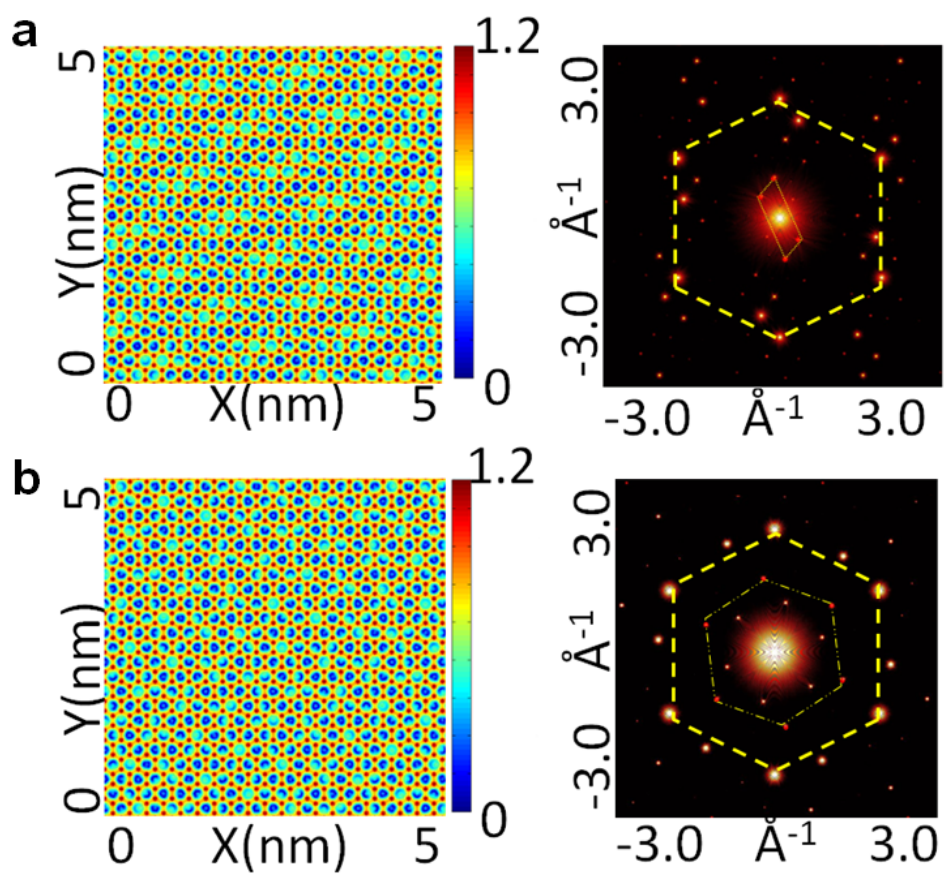

**Supplementary Figure 9.** Simulations of the real-space Moiré patterns (left panels) and the corresponding FT (right panels) for the 2D graphene honeycomb lattice on (a) Cu (100) square lattice and (b) Cu (111) hexagonal lattice. The FT Moiré pattern for graphene on Cu (100) is consistent with a honeycomb lattice at $\theta = (12\pm2)°$ relative to the square lattice of Cu (100), whereas that for graphene on Cu (111) is consistent with a honeycomb lattice at $\theta = (6\pm2)°$ relative to the hexagonal lattice of Cu (111). The FT Moiré patterns of **a,** and **b,** compare favorably with the FT spectra of Fig. 3b and Fig. 3c, respectively.

# Supplementary Notes:

## 1. Analysis of the plasma in PECVD growth

The optical emission spectra (OES) of the plasma were measured using a fiber coupled spectrometer (S2000, Ocean Optics, Inc) placed directly below the Evenson cavity. Absorption spectra were taken prior to graphene deposition away from the area of the copper sample. The spectra were referenced to a pure hydrogen plasma, *i.e.*, before the addition of the flow of methane. Shown in Supplementary Figure 2 is a typical emission spectrum featuring CN (388 nm), CH (431 nm), and CC (516 nm),[1] where negative absorption peaks indicate an increase in a component and positive peaks indicate a decrease.

Graphene growth was performed on a variety of copper substrates, including high purity copper foil, common OFHC sheet, single crystal (100), and single crystal (111). Gas chromatography was used to measure the amount of methane in the hydrogen stream during graphene growth. The gas chromatograph (GC) was an HP 5890 II employing a flame ionization detector (FID). The FID was calibrated using reference standards (Mesa Gass, Inc) of 500, 1000, 2000, 4000, and 10000 ppm of $CH_4$ in nitrogen. The response (integrated peak area) of the FID to methane was found to be linear and is shown in Supplementary Figure 3.

Based on our investigations, we believe that the most likely plasma species acting upon Cu during the PECVD graphene growth are atomic hydrogen and CN radicals. Atomic hydrogen via a hydrogen plasma is known to be effective for removing native atmospheric derived species including $Cu_2O$, $CuO$, $Cu(OH)_2$, and $CuCO_3$. It has been recently reported that CN radicals can be highly reactive towards removing Cu from a semiconductor at room temperature.[2] Our overall observation of the PECVD graphene growth was in agreement with this notion. As an example, we increased the amount of nitrogen present in the system slightly and found that the PECVD growth could occur at half the normal plasma power. Conversely, under excess methane conditions the Cu substrate would not etch at even more than double the normal plasma power. The presence of both atomic hydrogen and CN species in the plasma could allow simultaneous preparation of the copper surface and deposition of high quality graphene at reduced temperatures. However, more experimental work will be necessary to fully understand the role of each radical plays in the growth process even though reproducible PECVD-growth without any active heating has been made.

While the presence of trace nitrogen in the plasma was essential for successful PECVD growth, we note that nitrogen was not incorporated into our PECVD-graphene, as verified by the studies of x-ray photoemission spectroscopy (XPS) in Supplementary figure 4a, where no discernible N 1s core level binding energy at 398.1eV could be resolved for a sample of monolayer graphene on Cu foil. We further note that the two small peaks at 407.5 eV (Peak A) and 395.5 eV (Peak B) are associated with the Auger lines of the Cu substrate rather than the nitrogen 1s core level in the graphene sample. This realization was achieved by comparing the XPS spectrum of a bare Cu foil as the control sample. As demonstrated in Supplementary Figure 4b, the XPS data obtained from the control sample (pure Cu foil) over the same energy range (black curve) are essentially identical to that of our graphene sample (red curve). Moreover, when the XPS source was changed from Mg to Al, Peak A and Peak B both disappeared, indicating that they must be Auger lines of the underlying Cu substrate and that there was no nitrogen present in our graphene sample, because the nitrogen binding energy cannot be dependent on the x-ray energy.

## 2. Substrate influence on the Raman spectra of graphene

While the 2D-band FWHM of graphene may be used for analyzing whether a graphene sample consists of monolayer or multi-layers, possible changing properties of the underlying substrate could lead to complications in interpreting the Raman spectroscopic data. For instance, the 2D-band FWHM of our PECVD-grown graphene on Cu immediately after growth was ~ 29 cm$^{-1}$, which increased with time after growth and exposure to air, as exemplified in Supplementary Figure 6c-d. Similarly broadened FWHM of the 2D-band (Supplementary Figure 6e) was also observed in a commercial monolayer graphene sample on copper (from Graphene Supermarket) grown by the thermal CVD method. These findings are consistent with oxidation of the underlying Cu substrate, as discussed by Yin *et al.*[3]

The oxidation of the Cu substrate for our PECVD-graphene is likely the result of plasma-induced damages to the top side of Cu during the graphene growth. On the other hand, the quality of our PECVD-graphene remains intact despite the oxidation of the underlying Cu substrate, which has been verified by Raman spectroscopic studies of one of the PECVD-graphene samples transferred from Cu to SiO$_2$ after approximately one year of its growth. As demonstrated in Supplementary Figure 7a-d, a large spatial map of the sample over an area of (160×150) μm$^2$ reveals a mean FWHM = 28.8 cm$^{-1}$ for the 2D-band and a mean 2D/G intensity ratio ($I_{2D}/I_G$) = 2.7, both are consistent with predominantly monolayer graphene if we use either the criterion FWHM < 30 cm$^{-1}$ for the 2D-band as the figure of merit for monolayer graphene on SiO$_2$,[4] or the criterion ($I_{2D}/I_G$) > 1 for monolayer graphene on various substrates.[5,6,7,8] We further note the apparent correlation between the spatial maps of Supplementary Figure 7a-b, suggesting the consistency for determining monolayer graphene by using either the criterion for the FWHM of the 2D-band or that for the intensity ratio ($I_{2D}/I_G$).

## Supplementary References: